\documentclass[doublecol,linenumbers]{epl2} % for 2 columns style with line numbers
\usepackage{amsmath,amssymb}

\title{Observing a quantum phase transition by measuring a single spin}
\author{Manuel Gessner\inst{1} \and Michael Ramm\inst{2} \and Hartmut H\"affner\inst{2} \and Andreas Buchleitner\inst{1} \and Heinz-Peter Breuer\inst{1}}
\shortauthor{Manuel Gessner \etal}

\institute{                    
  \inst{1} Physikalisches Institut, Albert-Ludwigs-Universit\"at Freiburg, Hermann-Herder-Stra\ss e 3, 79104 Freiburg, Germany\\
  \inst{2} Department of Physics, University of California, Berkeley, California 94720, USA
}


\abstract{
We show that the ground state quantum correlations of an Ising model can be detected by monitoring the time evolution of a single spin alone, and that the critical point of a quantum phase transition is detected through a maximum of a suitably defined observable. A proposed implementation with trapped ions realizes an experimental probe of quantum phase transitions which is based on quantum correlations and scalable for large system sizes.}

\begin{document}

\maketitle

%\revision{Insert here the text.
%See fig.~\ref{fig.1}, table~\ref{tab.1} and eq.~(\ref{eq.1}).
%See also~\cite{b.a,b.b}.}

%\begin{figure}
%\onefigure{epl-template.eps}
%\caption{Figure caption.}
%\label{fig.1}
%\end{figure}

Classical phase transitions occur when a system reaches a critical temperature. A quantum phase transition occurs at zero temperature and describes a discontinuous qualitative change of the ground state when an external parameter is varied \cite{QPTReview}. Cold atoms and trapped ion systems represent ideal platforms for experimental simulations of complex quantum models \cite{CHRISTIAN,ColdAtoms,AbuChaosReview}, permitting studies of quantum phase transitions \cite{Markus,Toyoda,Islam11}. In particular, spin chain models of magnetic systems have been realized experimentally \cite{CHRISTIAN,Tobi,Islam11,Kim,Britton,nat1,nat2}. Signatures of an emerging quantum phase transition from a paramagnetic to a ferromagnetic phase have been observed in a trapped ion simulation of an Ising spin chain through measurements of the average magnetization \cite{Islam11}. 

The change of the ground state in the course of a quantum phase transition is reflected by the behavior of the quantum correlations, which therefore may also serve as an indicator of such a transition. Despite numerous theoretical studies \cite{Lidar,Fazio,Nielsen,Dillenschneider}, so far, there have been no direct experimental observations which connect this fundamental concept from quantum information theory to the theory of phase transitions. In trapped-ion simulations of quantum magnets, entanglement has only been explicitly studied between two or three spins \cite{Tobi,Kim}. This is due to the fact that the detection of quantum correlations usually requires simultaneous operations on some of the subsystems, involving a rapidly increasing experimental overhead for larger systems \cite{BlattWineland,Toth}.

These difficulties may be circumvented with the aid of a recently proposed local detection protocol \cite{GB,GB1}, which allows for the detection of quantum correlations by measuring only a small part of the system. This leads to a scalable detection scheme, applicable to systems which are too large for conventional methods. Recent experiments have demonstrated that this protocol can be implemented in systems of trapped ions \cite{NPhys} and photons \cite{Hefei} with current technology.

In this work, we apply the local detection protocol to a trapped-ion quantum simulation of the Ising model for long-range ferromagnetic interaction. The protocol is based on a dephasing operation which is implemented locally on one of the spins and erases all quantum correlations between the addressed spin and the rest of the chain. One compares the time evolution of the selected spin with and without application of the dephasing operation. Any difference of these evolutions indicates the presence of quantum correlations in the initial state. Moreover, it is possible to provide a lower bound on the initial quantum correlations by monitoring the time evolution of the trace distance of the spin states. We also discuss an extension of the protocol which involves dephasing in different bases and avoids overestimation of the total quantum correlations in case of a mixed state. The entire protocol is carried out by local operations and measurements on the single spin only, regardless of the total length of the chain. We show that the locally obtainable signal of the ground state quantum correlations attains a global maximum when the system approaches the critical point of a quantum phase transition. The signal is sufficiently strong to be observable under realistic experimental conditions and fairly robust against finite-temperature corrections.

The theoretical method presented here is not limited to one-dimensional spin chains, but rather applicable in a very general context \cite{GB}, including two- or three-dimensional ion crystals \cite{Britton}, provided access to one individual spin can be established. The method can also be applied when local control extends beyond two-dimensional observables \cite{GB}.

We now turn to discussion of the long-range Ising model with transverse field, described by the Hamiltonian
\begin{equation}
H = -\sum_{\substack{i,j=1\\(i < j)}}^N J_{ij} \sigma_x^{(i)} \sigma_x^{(j)} - B \sum_{i=1}^N \sigma_y^{(i)},
\end{equation}
where $\sigma_x^{(i)}$ and $\sigma_y^{(i)}$ are Pauli spin operators applied to the $i$th spin. For a quantum simulation one applies appropriate laser fields to a set of trapped ions, exerting a state-dependent optical dipole force \cite{PorrasCiracSpins}. This allows to engineer the spin-spin interactions via coupling to collective vibrations. An effective magnetic field $B$ is simulated by driving Rabi oscillations between the ion's electronic states. The laser-induced Ising coupling $J_{ij}$ approximately follows the scaling $J_{ij} \approx J_0/|i-j|^\alpha$, with $0 < \alpha < 3$ \cite{Kim}. For $J_0>0$ ($J_0<0$) the interaction is (anti-)ferromagnetic. From now on we restrict to the ferromagnetic case, but our analysis can be readily adapted to anti-ferromagnetic couplings. %The experimental procedure begins by initializing all the spins to point along the $y$ direction \cite{Tobi,Islam11}. This is the exact ground state of the Hamiltonian $H$ when $J_0=0$ and an approximate ground state for very strong B fields: $B / J_0\gg 1$. After the preparation step, the laser fields which generate the Ising couplings are turned on and a large effective magnetic field $B$ is applied \cite{Tobi,Islam11}. Then, the parameter $B/J_0$ is adiabatically ramped down, such that the system remains in the ground state of $H$ during the ramp \cite{Tobi,Islam11}.
Note that a finite $\alpha$ leads to long-range spin-spin couplings and excludes the possibility to obtain an exact analytical solution by employing the well-known Jordan-Wigner transformation \cite{JW}, which is limited to nearest-neighbor interactions.

In order to gain intuition for the system, we first consider the special case $J_0=0$ and $B>0$. The system's ground state is a pure product state in which all spins point along the $y$-direction. Conversely, for $J_0>0$ and $B = 0$ the two-fold degenerate ground state manifold is spanned by the ferromagnetic states $|\uparrow \uparrow \ldots \rangle $ and $|\downarrow \downarrow \ldots \rangle $, where $|\uparrow \rangle$ and $| \downarrow \rangle $ are the eigenstates of $\sigma_x$. As $B/J_0$ is increased, the competition between spin-spin interactions and the external magnetic field eventually leads to a transition from ferromagnetic to paramagnetic phases in the thermodynamical limit. The quantum phase transition can be observed in the spectrum: For small values of $B/J_0$, the ground state manifold is two-fold degenerate. This degeneracy is lifted when $B/J_0$ exceeds a critical value, whose exact position depends on $\alpha$ and  typically is found around $B\approx J_0$ for interactions of moderate range. In finite systems there is always an energy gap between the two lowest lying states, which approaches zero as $B/J_0$ decreases.% [Fig.~\ref{fig.spectrum}]. %Still, a clear signature of the quantum phase transition can already be observed in spin chains of less than 10 spins, e.g., by measuring the probability to have $s$ spins in the $|\uparrow\rangle$ state for $s=0,\dots,N$ \cite{Islam11}.

%\begin{figure}
%\includegraphics[width=.49\textwidth]{Spectrum.pdf}
%\caption{Lower part of the energy spectrum of the quantum Ising model for different numbers of spins and $\alpha=1$. In the thermodynamical limit, the gap between the two ground states (see also inset) is zero until the critical point of the quantum phase transition is reached and the degeneracy of the ground state manifold is lifted. A signature of the phase transition can be observed in finite systems and becomes sharper for increasing system size.}
%\label{fig.spectrum}
%\end{figure}

As pointed out by a number of theoretical studies, quantum correlations play an important role in the characterization of quantum phase transitions \cite{Lidar,Fazio,Nielsen,Dillenschneider}. Due to the $Z_2$ symmetry of the Hamiltonian $H$, all eigenstates have definite parity with respect to the $Z_2$ operation ($\sigma^{(i)}_x\rightarrow-\sigma^{(i)}_x$, $\sigma^{(i)}_y\rightarrow\sigma^{(i)}_y$, $\sigma^{(i)}_z\rightarrow-\sigma^{(i)}_z$). As a consequence, when $B/J_0$ is small, the ground state contains a superposition of Greenberger-Horne-Zeilinger type: $|\mathrm{GHZ}_{\varphi}\rangle=(|\uparrow \uparrow \ldots \rangle +e^{i\varphi}|\downarrow \downarrow \ldots \rangle)/\sqrt{2}$ \cite{Buzek,W}. These states are highly entangled across an arbitrary bipartite division, but show no quantum correlations when one or more spins are traced out \cite{W}. As $B/J_0$ is increased the spins begin to align along the magnetic field direction and eventually approach a product state in the limit $B/J_0\rightarrow\infty$. Hence, the bipartite global ground state entanglement decreases asymptotically towards zero.

\section{Estimating ground state negativity with single-spin measurements}In order to probe the quantum correlations of the ground state $|\Psi_0\rangle$ for different values of $B/J_0$ we employ the local detection protocol \cite{GB,GB1,NPhys}, restricting ourselves to local operations on the left-most spin of the chain. Note that if the total state is known to be pure, the single-spin entropy already reveals the degree of entanglement with the remaining system. However, the protocol does not rely on this assumption and an application to general mixed states is discussed later in the manuscript. 

We write the state $|\Psi_0\rangle$ in its Schmidt decomposition across a bipartite division between the first spin and the rest of the chain as $|\Psi_0\rangle=\sum_i\lambda_i|\varphi_i\rangle\otimes|\chi_i\rangle$ with Schmidt coefficients $\lambda_i$. The first step of the local detection protocol consists in performing a state tomography of the reduced density matrix $\rho_S$ of this spin, which is obtained by tracing over the remaining spins, $\rho_S=\mathrm{Tr}_R|\Psi_0\rangle\langle\Psi_0|$. When the reduced state $\rho_S$ does not contain degenerate eigenvalues, its eigenbasis is given by the local Schmidt basis $\{|\varphi_i\rangle\}_{i=0,1}$. Using this basis, a local dephasing operation on the full ground state is defined by
\begin{align}
\Phi(|\Psi_0\rangle\langle\Psi_0|)=\sum_{i=0,1}\left(|\varphi_i\rangle\langle\varphi_i|\otimes\mathbb{I}\right)|\Psi_0\rangle\langle\Psi_0|\left(|\varphi_i\rangle\langle \varphi_i|\otimes\mathbb{I}\right),
\end{align}
where $|\varphi_i\rangle\langle \varphi_i|$ operates on the leftmost spin and $\mathbb{I}$ denotes the identity operation on the rest of the chain. This corresponds to a non-selective local measurement, which leads to complete dephasing in the single spin's eigenbasis. As demonstrated in ref.~\cite{NPhys}, this operation can be experimentally implemented on a trapped ion with a laser-induced AC-Stark shift and additional unitary single-spin operations. The effect of this operation is to convert a coherent superposition into the corresponding incoherent mixture, thereby removing the quantum correlations of the original state $|\Psi_0\rangle$. The ground state entanglement can be quantified by the trace distance of the original ground state $|\Psi_0\rangle\langle\Psi_0|$ and the locally dephased state $\Phi(|\Psi_0\rangle\langle\Psi_0|)$ where the quantum correlations have been removed. Since the measured subsystem is a qubit, one can show, employing the Schmidt decomposition, that
\begin{align}\label{eq.totaltracedist}
D(|\Psi_0\rangle\langle\Psi_0|)&:=\frac{1}{2}\left\||\Psi_0\rangle\langle\Psi_0|-\Phi(|\Psi_0\rangle\langle\Psi_0|)\right\|\notag\\&=\mathcal{N}(|\Psi_0\rangle\langle\Psi_0|),
\end{align}
where $\mathcal{N}(\rho)=(\|\rho^{\Gamma}\|-1)/2$ is the negativity, a well-known entanglement measure \cite{VidalWerner}. We have introduced the trace norm $\|X\|=\mathrm{Tr}\sqrt{X^{\dagger}X}$ \cite{NielsenChuang} and $\rho^{\Gamma}$ denotes the state $\rho$ after partial transposition in one of the subsystems in an arbitrary basis. Eq.~(\ref{eq.totaltracedist}) holds generally as long as the total state is pure and the measured system is a qubit, which is locally dephased in the Schmidt basis. 

\begin{figure}[tb]
\includegraphics[width=.49\textwidth]{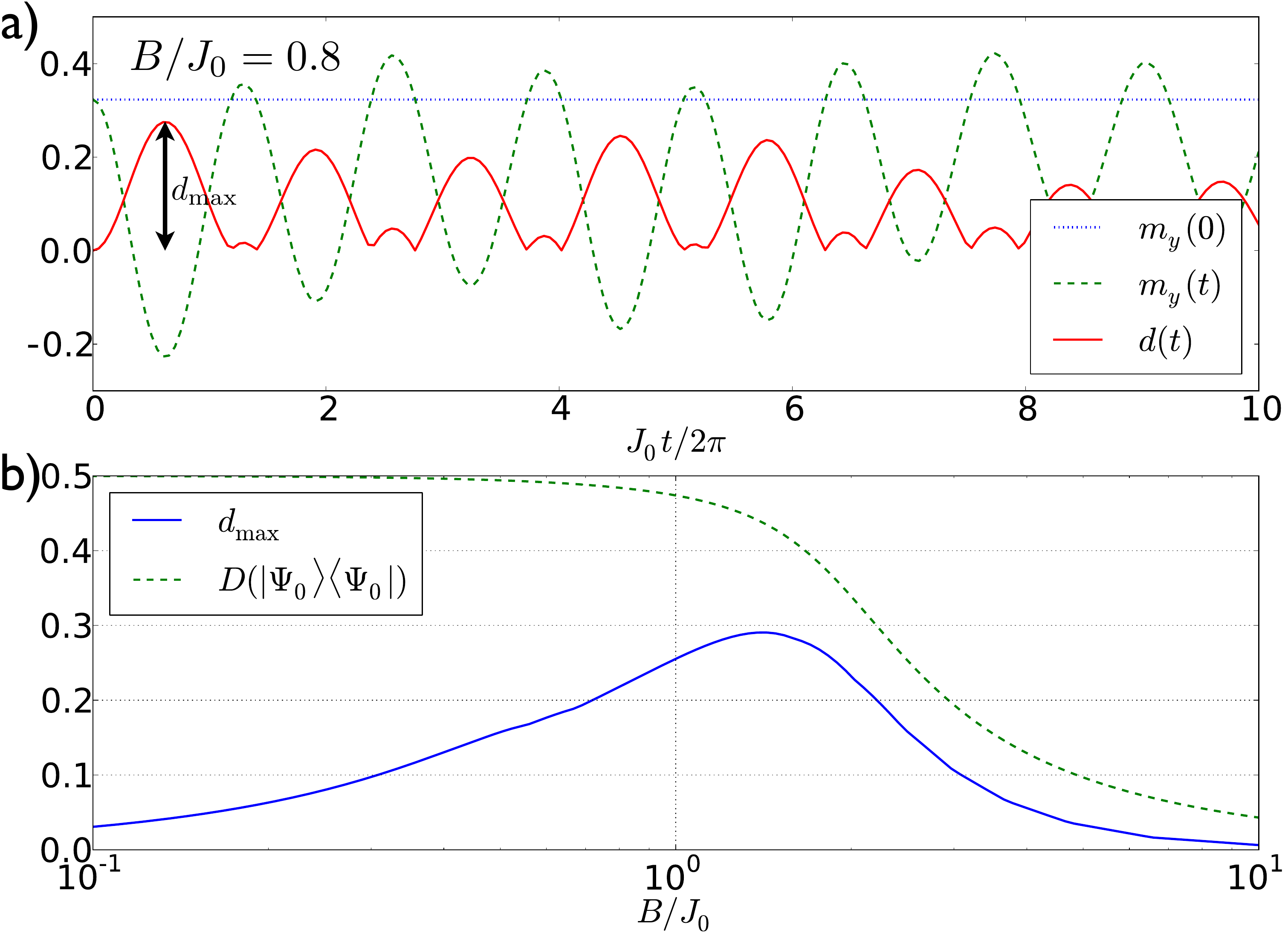}
\caption{a) The evolution of the locally dephased ground state can be observed by measuring the single-spin magnetization $m_y(t)$. The initial value $m_y(0)$ coincides with the time-invariant magnetization of the ground state before dephasing. The local trace distance between ground state and dephased state is given by $d(t)=1/2|m_y(t)-m_y(0)|$. The plot shows the time evolution for a chain of $7$ spins for the experimentally observable time range of $t\leq2\pi\times10/J_0\approx 5\,\mathrm{ms}$. The maximum value $d_{\mathrm{max}}$ represents a witness for the total correlations. b) The ground state spin-rest entanglement $D(|\Psi_0\rangle\langle\Psi_0|)$ decreases from its maximum value to zero when $B/J_0$ is increased. The local witness $d_{\mathrm{max}}$ reaches a maximum close to the critical point of the quantum phase transition.% By scanning out further, the signal can be improved and $d_{\mathrm{max}}$ gets smoothened. This can be seen by the red dotted line, showing $d_{\mathrm{max}}$ when maximized over an extended time interval of $50\,\mathrm{ms}$.
}
\label{fig.correlationswitness}
\end{figure}

While the ground state $|\Psi_0\rangle$ does not evolve in time under the action of the Hamiltonian $H$, the application of the dephasing operation to the ground state typically does not produce an eigenstate of the Hamiltonian $H$. This state will evolve as $U(t)\Phi(|\Psi_0\rangle\langle\Psi_0|)U^{\dagger}(t)$ with $U(t)=e^{-iHt}$. The reduced dynamics of the single spin,
$\rho_S(t)=\mathrm{Tr}_R\{U(t)\Phi(|\Psi_0\rangle\langle\Psi_0|)U^{\dagger}(t)\},$ 
can be observed by single-spin measurements at different times $t$. Since the dephasing $\Phi$ is performed in the eigenbasis of the initial reduced state, we obtain $\rho_S(0)=\mathrm{Tr}_R\{\Phi(|\Psi_0\rangle\langle\Psi_0|)\}=\mathrm{Tr}_R\{|\Psi_0\rangle\langle\Psi_0|\}$ \cite{GB}. Note that also the reduced state of the remaining spin chain remains unaffected by the dephasing due to its local nature \cite{GB}. By comparing the deviation of the state $\rho_S(t)$ from the initial state $\rho_S(0)$ we can estimate the quantum correlations in the initial state $|\Psi_0\rangle$ [fig.~\ref{fig.correlationswitness} a)]. More precisely, for every $t$, the trace distance
\begin{align}
d(t)=\frac{1}{2}\|\rho_S(t)-\rho_S(0)\|
\end{align}
provides a locally accessible lower bound for the total correlations $D(|\Psi_0\rangle\langle\Psi_0|)$ \cite{GB1}. This is a consequence of the contractivity of the trace distance under positive operations, such as the time evolution and the partial trace \cite{NielsenChuang}. The optimal bound is found by maximizing over all measurement times $t$, leading to \cite{NPhys}
\begin{align}\label{eq.maxdist}
d_{\mathrm{max}}=\max\limits_{t}d(t)\leq D(|\Psi_0\rangle\langle\Psi_0|).
\end{align}
Since $\rho_S(t)$ is always diagonal in the $\sigma_y$ basis, i.e., $\rho_S(t)=1/2(\mathbb{I}_2+m_y(t)\sigma_y)$ with $m_y(t)=\mathrm{Tr}\{\rho_S(t)\sigma_y\}$, the trace distance $d(t)$ is fully determined by the magnetization along $y$ as $d(t)=1/2|m_y(t)-m_y(0)|$, see fig.~\ref{fig.correlationswitness} a).

Fig.~\ref{fig.correlationswitness} b) shows the ground state negativity $D(|\Psi_0\rangle\langle\Psi_0|)$ between the measured spin and the rest of the chain and its local signature $d_{\mathrm{max}}$ as a function of $B/J_0$. In all of our simulations we use $\alpha=1$. For very small values of $B/J_0$, the contributions of $|\mathrm{GHZ}_{\varphi}\rangle$ to the ground state lead to strong bipartite entanglement (green dashed line) \cite{Remark}. The local signal (continuous blue line) is generated by the time evolution of the dephased state, which becomes richer close to the critical point and leads to a stronger signal. For very large $B/J_0$ the ground state $|\Psi_0\rangle$ approaches the completely uncorrelated product state $|\Psi_0\rangle=|\uparrow_y\rangle^{\otimes N}$, where all spins point along the direction of the magnetic field. Since this state is separable, it is invariant under the action of the dephasing. The local signal thus reaches a peak as the magnetic field strength approaches the critical point of the phase transition. Hence, the detected quantum correlations unveil a clear signature of the quantum phase transition, which is experimentally accessible with operations and measurements on a single spin only. Since the signal is based on the local magnetization, which is not reduced when the number of spins increases, we expect to observe qualitatively similar signals also for larger $N$. 

The local signal is created by two processes: The removal of quantum correlations by dephasing, and the subsequent single-spin dynamics of the thereby populated excited states. By only considering a finite time-window, and tracing over all other remaining spins, some information about the initial quantum correlations is lost. Thus, generally, it is difficult to exactly determine the critical point based on the local signal. Note, however, that only in the thermodynamical limit and for $\alpha=\infty$, the critical point is found at $B/J_0=1$. Finite values of $\alpha$ shift the critical point to larger values and in the limit $\alpha\rightarrow 0$ it moves towards $B/J_0\rightarrow\infty$.

The parameters used for the simulations \cite{qutip} are based on the parameters of the experiments reported in refs.~\cite{Tobi,Islam11,Kim}. Typical parameters are $J_0\approx 2\pi\times 500\,\mathrm{Hz}$, with a coherence time of approximately $3\,\mathrm{ms}$. We thus restrict the time window for the optimization of $d(t)$ to a realistic range of $5\,\mathrm{ms}$ with a resolution of $200$ timesteps. %This explains the dips which can be observed in the signal, especially for larger number of spins. 
By considering a longer time window for optimization in our simulations, the lower bound can be further improved.

The experimental procedure consists of the following steps: First, the ground state $|\Psi_0\rangle$ is prepared, e.g., by adiabatically ramping down the effective $B$-field after initializing all spins along the $y$ direction. After obtaining the single-spin density matrix $\rho_S(0)$, the process is repeated and the dephasing is implemented. By monitoring the time evolution of the single spin after dephasing, $\rho_S(t)$, and comparing its deviation from the ground state, one obtains the local trace distance $d(t)$.

\section{Extension to mixed states and finite-temperature corrections}In the thermodynamical limit the energy gap $\Delta$ between the two energetically lowest states is exactly zero for values of $B/J_0\lesssim 1$ and we will refer to these two states as the two ground states even if they are not degenerate. For finite systems $\Delta$ tends towards zero for small $B/J_0$. Therefore, instead of the pure ground state, it is more realistic to consider a low-temperature thermal state $\rho_{\beta}=e^{-\beta H}/\mathrm{Tr}e^{-\beta H}$ with $\beta=1/kT$.

We now discuss application of the local detection protocol to a mixed state, and develop an extension of the protocol to avoid overestimation of the quantum correlations. Dephasing in the local eigenbasis $\{|\varphi_i\rangle\}_{i=0,1}$ of $\rho_S=\mathrm{Tr}_R(\rho)$ always produces states of the form $\Phi(\rho)=\sum_{i=0,1}p_i|\varphi_i\rangle\langle\varphi_i|\otimes\rho_i^R$, which are known as classically correlated states \cite{Review}. The reduced density matrices for both, the selected spin and the rest of the chain, coincide for $\rho$ and $\Phi(\rho)$, which means that in the local subsystems the two states cannot be distinguished. One can therefore interpret $\Phi(\rho)$ as the corresponding classically correlated counterpart to $\rho$ \cite{GB1}. The trace distance $D(\rho)=\left\|\rho-\Phi(\rho)\right\|/2$ quantifies their distinguishability in the total system \cite{NielsenChuang} and is related to the concept of measurement-induced disturbance \cite{Luo}. %Degenerate eigenvalues in $\rho_S(0)$ pose a problem for the definition of $D(\rho)$ since the eigenbasis is not uniquely defined in this case \cite{GB1}. Otherwise this measure is faithful in a sense that it is zero if and only if the state $\rho$ is classically correlated. However 
For certain mixed total states, $D$ is known to overestimate the amount of quantum correlations \cite{AMID,Campbell}, which typically are defined in terms of the minimum over all measurement bases \cite{Review}. It is therefore necessary to introduce the minimal measurement disturbance
\begin{align}\label{eq.mindist}
D_{\mathrm{min}}(\rho)=\min_{\Pi}D_{\Pi}(\rho)=\min_{\Pi}\frac{1}{2}\left\|\rho-\Phi_{\Pi}(\rho)\right\|,
\end{align}
where $\Phi_{\Pi}(\rho)=\sum_{i=0,1}(\Pi_i\otimes\mathbb{I})\rho(\Pi_i\otimes\mathbb{I})$ describes local dephasing in a basis characterized by the two orthogonal projectors $\Pi=\{\Pi_0,\Pi_1\}$. 
%The set of dephasing operations includes $\Phi$, i.e., dephasing in the Schmidt basis of $\rho_S$, as a special case when $\Pi=\{|\varphi_0\rangle\langle\varphi_0|,|\varphi_1\rangle\langle\varphi_1|\}$. 
When the measured system is a qubit, it can be shown that $D_{\mathrm{min}}(\rho)$ coincides with the minimum entanglement potential \cite{NoQ}, a measure for the nonclassical correlations which can be activated into entanglement with an ancilla system \cite{Piani,Bruss}. For pure states, the minimum entanglement potential again equals the conventional negativity $\mathcal{N}$ \cite{QoC}. Combined with eq.~(\ref{eq.totaltracedist}) this implies that for pure states the basis which minimizes the trace distance of dephased and original state [eq.~(\ref{eq.mindist})] is given by the local Schmidt basis, which is locally accessible.

\begin{figure}[tb]
\includegraphics[width=.49\textwidth]{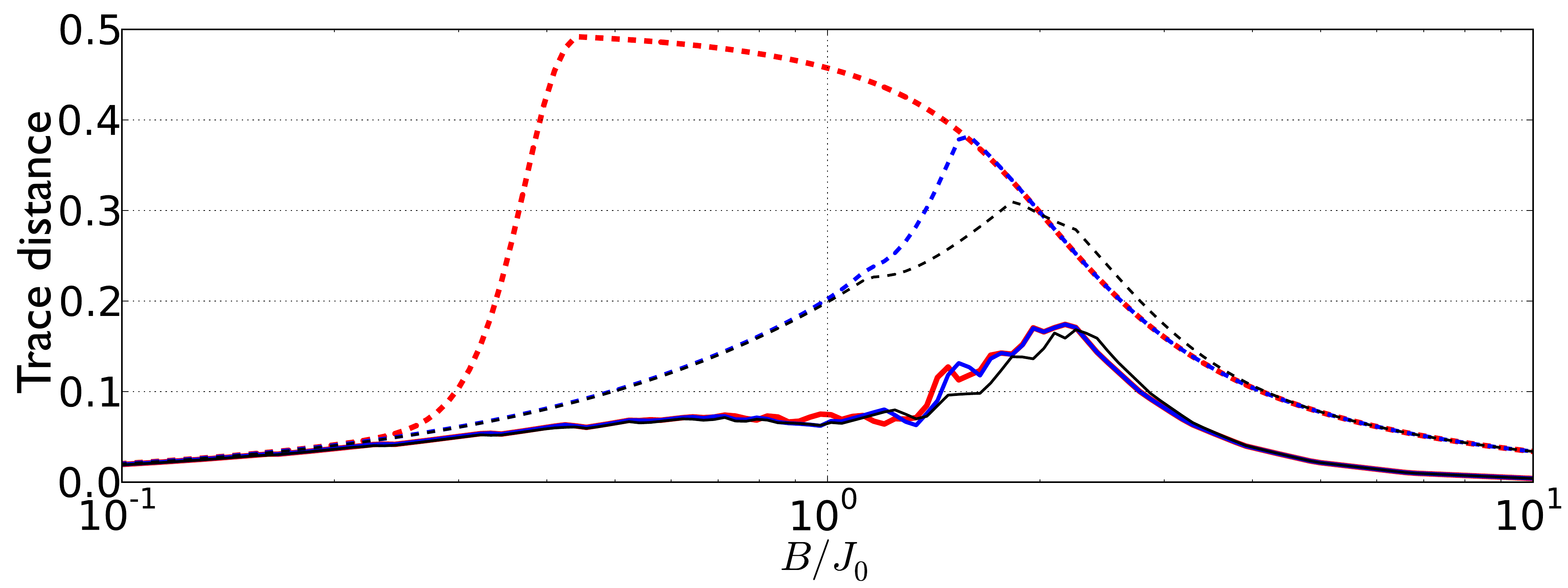}
\caption{The thermal state $\rho_{\beta}$ is mostly determined by the ground state until the size of the energy gap $\Delta$ between the two energetically lowest states becomes comparable to the thermal energy $\beta^{-1}=kT$. Then the state becomes an incoherent mixture of the lowest lying states which washes out the quantum correlations $D_{\mathrm{min}}(\rho_{\beta})$ (dashed lines). The single-spin signal $d_{\mathrm{min}}(\rho_{\beta})$ (continuous lines) is barely affected by finite temperature effects. In the plot we have $N=7$ spins and $kT/J_0 = 10^{-5}$ (thick, red), $0.1$ (medium, blue), and $1.0$ (thin, black). The optimization over dephasing measurements was done over a sample of 20 uniformly distributed pairs of basis states.}
\label{fig.amid}
\end{figure}

For $kT \ll \Delta$, i.e., when the temperature is much lower than the energy gap between the two ground states, $\rho_{\beta}$ corresponds to the energetically lower pure ground state $|\Psi_0\rangle$. However for $kT \approx \Delta$, the thermal state $\rho_{\beta}$ approaches the equally-weighted incoherent mixture of the two ground states (see also \cite{Nielsen}). Here we assume that the first excited state is energetically separated from the two ground states by much more than $kT$, which is the case for small $B/J_0$. An equal mixture of $|\mathrm{GHZ}_{\varphi}\rangle$ and $|\mathrm{GHZ}_{\varphi+\pi}\rangle$ contains no quantum correlations. As a consequence, the thermal quantum correlations drop from their maximal value to zero when $B/J_0$ falls below a value where $\Delta \approx kT$. This is reflected by $D_{\mathrm{min}}(\rho_{\beta})$ in fig.~\ref{fig.amid}.%, whereas $D(\rho_{\beta})$ shows the same behavior as $D(|\Psi_0\rangle\langle\Psi_0|)$ as long as $kT$ is smaller than the energy gap from the ground states to the first excited states.

In order to find a locally accessible lower bound for $D_{\mathrm{min}}(\rho_{\beta})$, we apply dephasing in a set of different bases and observe the local time evolution 
\begin{align}
d_{\Pi}(t)=\frac{1}{2}\left\|\rho_S(0)-\rho^{\Pi}_S(t)\right\|,
\end{align}
where $\rho_S(0)=\mathrm{Tr}_R\rho_{\beta}$ and $\rho^{\Pi}_S(t)=\mathrm{Tr}_R\{U(t)\Phi_{\Pi}(\rho_{\beta})U^{\dagger}(t)\}$. Note that unless $\Pi$ contains the projectors onto eigenstates of $\rho_S(0)$, the initial value $d_{\Pi}(0)$ is not necessarily zero. A lower bound for $D_{\mathrm{min}}(\rho)$ is given by the minimal local witness
\begin{align}\label{eq.optimizedwitness}
d_{\mathrm{min}}(\rho_{\beta})=\max_t\min_{\Pi}d_{\Pi}(t) \leq D_{\mathrm{min}}(\rho_{\beta}).
\end{align}
%This means that for every value $t$, one must pick the minimum of $d_{\Pi}(t)$. All of these minima represent a lower bound for $D_{\mathrm{min}}(\rho)$. The optimal bound is found again by the maximum over all values of $t$. 
Even if the total state is almost pure, the obtained bound may be lower than $d_{\mathrm{max}}$ since the minimum over $d_{\Pi}(t)$ may be attained by a different projector than the one which minimizes $D_{\Pi}(\rho_{\beta})$. %However, since $d_{\mathrm{min}}(\rho_{\beta})$ provides a bound for $D_{\mathrm{min}}(\rho)$, its information-theoretic interpretation is meaningful for arbitrary total states.

In fig.~\ref{fig.amid} we compare $d_{\mathrm{min}}(\rho_{\beta})$ and $D_{\mathrm{min}}(\rho_{\beta})$ for different values of $B/J_0$. We find that the minimization barely affects the local signal $d_{\mathrm{min}}(\rho_{\beta})$. The minimal local witness still reveals the critical point of the quantum phase transition and is more robust against finite temperature corrections than the total correlations $D_{\mathrm{min}}(\rho_{\beta})$.

%Pairs of orthogonal projectors can be characterized by $\Pi_{\pm}=1/2(\mathbb{I}\pm\vec{v}\cdot\vec{\sigma})$, where $\vec{\sigma}=(\sigma_x,\sigma_y,\sigma_z)$ is a vector of Pauli matrices and $\vec{v}=(\sin(\theta)\cos(\phi),\sin(\theta)\sin(\phi),\cos(\theta))$ is a unit vector in $\mathbb{R}^3$, which in turn is determined by the angles $\theta\in[0,\pi/2]$ and $\phi\in[0,\pi)$. % In Fig.~\ref{fig.} we sample over 5 equally spaced values of $\theta$ and 4 values of $\phi$ within the given ranges, leading to a total sample of 20 pairs of orthogonal projectors.

%\section{Conclusions}
In conclusion, we have shown that the quantum correlations in an Ising model can be detected with the aid of a local dephasing operation on a single spin and the subsequent observation of the spin's time evolution. The detection protocol does not rely on controlling other spins in the chain. This method provides a lower bound on the bipartite quantum correlations between the selected spin and the rest of the chain, which for pure states coincides with the negativity. When additionally dephasing is implemented over a complete set of projectors rather than only the local eigenbasis, the witness can be further refined such as to avoid overestimation of the quantum correlations in a mixed state. %We further obtained a general result for pure total states subject to local measurements on a qubit subsystem: The basis which minimizes the trace distance between original and locally measured state is given by the Schmidt basis of the qubit.

Simulations based on realistic parameters indicate that the attainable lower bound for ground state quantum correlations in the Ising model is large enough to be detected in state-of-the-art experimental quantum simulations of spin chains. When the effective magnetic field is tuned, the single-spin signature of the ground state quantum correlations assumes its maximum value at the critical point of the quantum phase transition. This signature is robust against finite-temperature corrections. Hence, following our proposal, we expect that an experimental detection of a quantum phase transition based on bipartite quantum correlations can be realized readily with existing setups, even for large chains, where other methods are not applicable. This constitutes a detection of a genuine many-body effect on the basis of single-particle operations.

The methods presented here can be applied readily to arbitrary spin-chain models, and are able to probe also higher-dimensional systems. Since the connection of quantum phase transitions to quantum correlations is not limited to this particular model -- in fact, it constitutes a rather generic effect --, we expect the presented method to be generally suitable to detect such phenomena in a wide range of scenarios. 

Finally we remark that quenching such a quantum system, e.g., by scanning the parameter $B/J_0$ through the phase transition towards zero within a certain quench time $\tau$, will induce defects whose density is proportional to the correlation length $1/\sqrt{\tau}$ \cite{quenches}. The bipartite entanglement between two blocks of spins was shown to scale with the same correlation length \cite{Cincio}, and can be detected with limited resources using the present method. This could be useful for an indirect entanglement-based verification of the universal scaling behavior of defect formation processes, which are of interest in various fields of physics \cite{Kibble,Zurek,quenches}.

\acknowledgments
We would like to thank G. Adesso, S. Wi\ss mann, S.A. Parameswaran and P. Richerme for discussions. M.G. thanks the German National Academic Foundation for support. This work was supported by the NSF CAREER Program Grant No. PHY 0955650.


\begin{thebibliography}{0}
\bibitem{Remark} From our numerical simulations for small $B/J_0$ we find $|\Psi_0\rangle\simeq|\mathrm{GHZ}_{\varphi}\rangle$ with $\varphi = N\pi/2$ leading to even parity with respect to the $Z_2$ operation, whereas for the energetically slightly higher state we have $|\Psi_1\rangle\simeq|\mathrm{GHZ}_{\varphi+\pi}\rangle$ with odd parity.
\end{thebibliography}
\end{document}